# THE EULER LEGACY TO MODERN PHYSICS

G. DATTOLI

ENEA - Dipartimento Tecnologie Fisiche e Nuovi Materiali
Centro Ricerche Frascati, Roma

M. DEL FRANCO

ENEA Guest

# THE EULER LEGACY TO MODERN PHYSICS

G. DATTOLI

ENEA -Dipartimento Tecnologie Fisiche e Nuovi Materiali
Centro Ricerche Frascati

M. DEL FRANCO - ENEA Guest

## THE EULER LEGACY TO MODERN PHYSICS

***Abstract***

*Particular families of special functions, conceived as purely mathematical devices between the end of XVIII and the beginning of XIX centuries, have played a crucial role in the development of many aspects of modern Physics.*
*This is indeed the case of the Euler gamma function, which has been one of the key elements paving the way to string theories, furthermore the Euler-Riemann Zeta function has played a decisive role in the development of renormalization theories.*
*The ideas of Euler and later those of Riemann, Ramanujan and of other, less popular, mathematicians have therefore provided the mathematical apparatus ideally suited to explore, and eventually solve, problems of fundamental importance in modern Physics.*
*The mathematical foundations of the theory of renormalization trace back to the work on divergent series by Euler and by mathematicians of two centuries ago.*
*Feynman, Dyson, Schwinger… rediscovered most of these mathematical "curiosities" and were able to develop a new and powerful way of looking at physical phenomena.*

**Riassunto**

Alcune particolari famiglie di funzioni speciali, concepite come dispositivi puramente matematici tra la fine del XVIII e l'inizio del XIX secolo, hanno svolto un ruolo cruciale nello sviluppo di molti aspetti della fisica moderna.
Questo è davvero il caso della funzione gamma di Eulero, che è stata una degli elementi chiave per aprire la strada alle teorie delle stringhe, inoltre, la funzione zeta di Eulero-Riemann ha svolto un ruolo decisivo nello sviluppo delle teorie di rinormalizzazione.
Le idee di Eulero e dopo quelle di Riemann, Ramanujan e di altri, meno popolari, matematici hanno quindi fornito gli apparati matematici ideali per esplorare, ed eventualmente risolvere, i problemi di fondamentale importanza nella fisica moderna.
I fondameti matematici della teoria di rinormalizzazione risalgono al lavoro sulle serie divergenti di Eulero e dei matematici di due secoli fa. Feynman, Dyson, Schwinger…riscoprirono molte di queste "curiosità" matematiche e furono capaci di sviluppare un nuovo e potente modo di osservare i fenomeni fisici.

**INDICE**

# THE EULER LEGACY TO MODERN PHYSICS

## 1. INTRODUCTION

Leonard Euler is universally recognized as one of the greatest mathematicians of all times. Albeit somebody believes him the greatest, the choice is however limited to him and Gauss and, just to make a comparison with great musicians, we can use the following equivalence: Euler:Mozart=Gauss:Beethoven.

The work and life of Euler can be characterized by a simple statement, which has the flavour of an oximoron: Genius and Regularity.

More than other comments his "natural" greatness was recognized by Condorcet, who, commemorating him, stated "… il cessa de vivre et de calculer". The calculus was as natural for him as for the other human beings breathing or for eagles flying.

The Euler's contribution to any branches of science have been seminal and any problem he touched has opened new fields of research in pure Mathematics or Physics. A good example is provided by the solution he gave to the popular problem of the seven bridges of Könisberg. The method proposed by him opened an entire new field of mathematics, known as Graph theory.

His great attitude was perhaps that of giving a meaning to notions which, at first sight, could be defined absurdities. He correctly handled the theory of imaginary numbers, introduced the concepts of fractional derivative and defined the Gamma function, which is the extension to non integers of the operation of factorial.

In this paper we will describe the relevance of Euler heritage to modern science and in particular to Physics. Our starting point will be Gamma function which, as a already stressed, was initially introduced to extend the operation of factorial to any real or complex number.

The definition proposed by Euler in 1729, in a letter to his friend Christian Goldbach, is reported below [1]

$$\Gamma(x) = \int_0^1 [-\ln(\sigma)]^{x-1} d\sigma, \qquad Re(x) > 0 \tag{1a}$$

After setting $\sigma = e^{-t}$, eq. (1a) can be rewritten in the more familiar way

$$\Gamma(x) = \int_0^\infty e^{-t} t^{x-1} dt, \qquad Re(x) > 0 \tag{1b}$$

An alternative form, yielding the Gamma in terms of the Gaussian function, is

$$\Gamma(x) = 2\int_0^\infty e^{-t^2} t^{2x-1} dt, \qquad Re(x) > 0 \tag{2}$$

The extension to non positive values of the real part of the variable $x$, was proposed by Euler himself (1729, letter to Goldbach) and later by Gauss in 1811 [1, 2]. The definition, valid without any restriction on the real or imaginary part of the variable $x$, is

$$\Gamma(x) = \lim_{p\to\infty} \Gamma_p(x), \qquad \Gamma_p(x) = \frac{p^x}{\prod_{n=0}^{p}\left(x + \frac{1}{n}\right)} \tag{3}$$

It ensures that the function $\Gamma(x)$ has an infinite number of poles, located at $x = 0, -1, -2, -3, \ldots$ (see Fig. 1 for the relevant plot, drawn on the real axis and in the complex plane).

An important step further in the theory of gamma function was due to Weierstrass [3], who proposed an infinite product formula, which determined significant successive developments.

By setting indeed

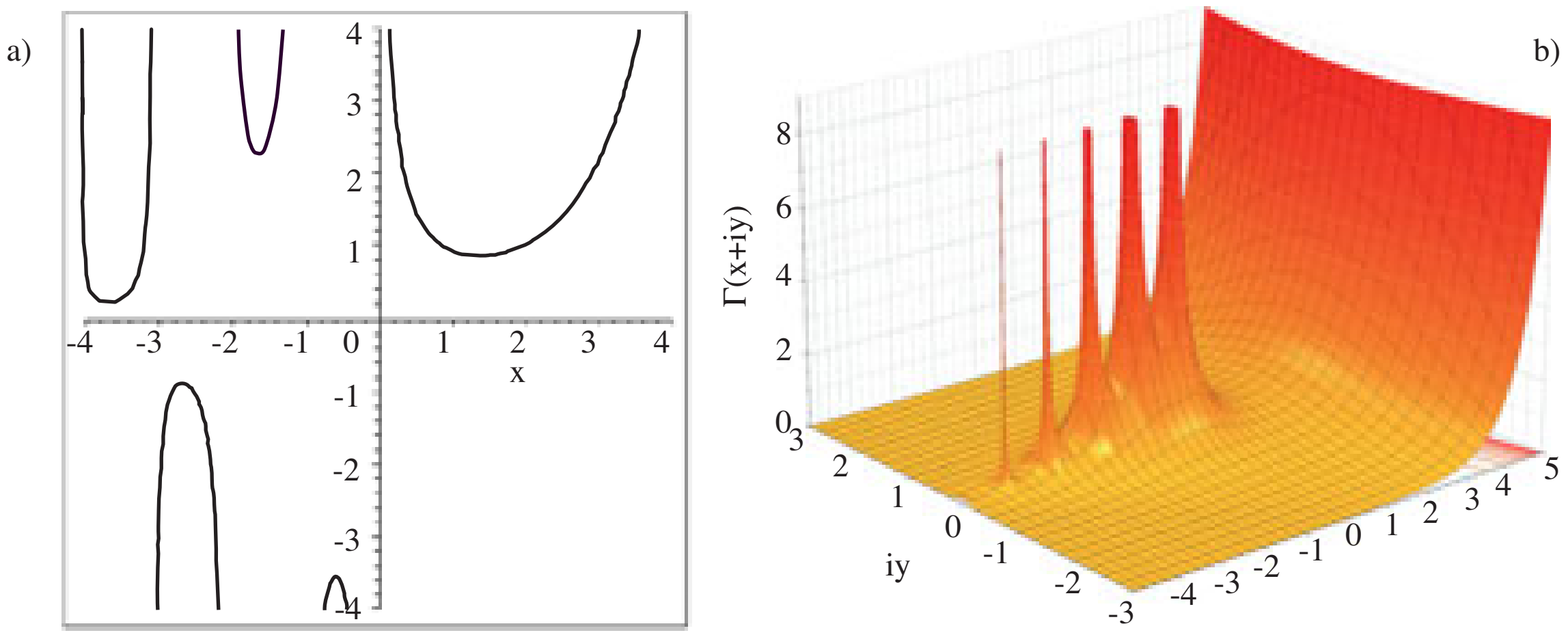

*Fig. 1 - Plot of the gamma function and of its modulus for a complex argument*

$$p^x = e^{x\,ln(p)} = e^{x\left[ln(p) - \sum_{s=1}^{p}\frac{1}{n}\right]} e^{x\sum_{n=1}^{p}\frac{1}{n}} \tag{4},$$

in eq. (3), we find

$$\frac{1}{\Gamma(x)} = x\,e^{\gamma\,x}\prod_{n=1}^{\infty}\left(1+\frac{x}{p}\right)\ e^{-\frac{x}{p}} \tag{5}$$

Where $\gamma$ is the Euler-Mascheroni constant, given by

$$\gamma = lim_{n\to\infty}\left[\sum_{s=1}^{n}\frac{1}{n} - ln(n)\right] \tag{6}$$

From the eq. (6) it also follows that

$$\frac{1}{\Gamma(x)\Gamma(-x)} = -x^2\prod_{n=1}^{\infty}\left(1-\frac{x^2}{n^2}\right) \tag{7a}$$

which, on account of the identities

$$sin(\pi x) = \pi x \prod_{n=1}^{\infty}\left(1-\frac{x^2}{n^2}\right),$$
$$\Gamma(-x) = -\frac{\Gamma(1-x)}{x} \qquad (7b),$$

yields the celebrated inversion formula

$$\Gamma(x)\Gamma(1-x) = \frac{\pi}{sin(\pi x)} \qquad (8).$$

The Gamma function has other important properties, which are reported below for the sake of completeness [3].

***Expansion around a singularity***

$$\Gamma(-n+\varepsilon) \cong \frac{(-1)^n}{n!}\left[\frac{1}{\varepsilon}+\psi(n+1)+\frac{\varepsilon}{2}\left(\frac{\pi^2}{3}+\psi(n+1)^2-\psi'(n+1)\right)\right]+o(\varepsilon^2),$$
$$\psi(z) = \frac{\Gamma'(z)}{\Gamma(z)} \qquad (9)$$

***The duplication theorem (Legendre (1809))***

$$\Gamma(x)\Gamma(x+\frac{1}{2}) = \frac{\sqrt{\pi}}{2^{2x-1}}\Gamma(2x) \qquad (10)$$

***Stirling formula***

$$lim_{x\to\infty}\,\Gamma(x) \cong \sqrt{2\pi x}\, x^x e^{-x}\left(1+o\left(\frac{1}{x}\right)\right) \qquad (11)$$

***Gauss multiplication theorem***

$$\Gamma(nx) = \frac{n^{nx}}{\sqrt{(2\pi)^{n-1}\,n}}\prod_{r=0}^{n-1}\Gamma\left(x+\frac{r}{n}\right) \qquad (12)$$

Which for $x = 1/n$ yields the Euler product

$$\prod_{r=0}^{n-1}\Gamma\left(\frac{r}{n}\right)=\frac{\sqrt{(2\pi)^{n-1}}}{n} \qquad (13).$$

Let us now consider the following integral

$$I(x,y)=\int_0^{\infty}e^{-(x-1)\xi}(1-e^{-\xi})^{y-1}d\xi \qquad (14)$$

representing a two variable generalization of the gamma function. After the transformation of variable $e^{-\xi}=t$, it can be cast in the form

$$\int_0^1 t^{x-1}(1-t)^{y-1}dt=B(x,y) \qquad (15).$$

The last integral provides the more familiar form of the Euler $B$ function, which can be considered the extension, to the non integers, of the binomial function. A straightforward manipulation shows indeed that

$$B(x,y)=\frac{\Gamma(x)\Gamma(y)}{\Gamma(x+y)}=\frac{1}{x}\binom{x+y-1}{y-1}^{-1} \qquad (16).$$

The $B$ function is evidently symmetric in the $x$ and $y$ variables and satisfies the further property

$$B(x,y)=\sum_{n=0}^{\infty}\frac{(-1)^n}{n!}\frac{1}{(x+n)}\frac{\Gamma(y)}{\Gamma(y-n)}=\sum_{m=0}^{\infty}\frac{(-1)^m}{m!}\frac{1}{(y+m)}\frac{\Gamma(x)}{\Gamma(x-m)} \qquad (17)$$

In the hand of Euler the functions, named after him, became a powerful tool to make tremendous progresses in any branches of pure and applied Mathematics. We report in the following two examples, stressing the importance of Gamma functions in the Calculus. In the forthcoming part of the paper we will discuss the importance in different branches of Physics.

***i) The formalism of derivatives of fractional order [4]***

Euler noted indeed that ordinary derivatives satisfy the relationship

$$\left(\frac{d}{dx}\right)^m x^n = \frac{n!}{(n-m)!} x^{n-m},$$
$$m > n \tag{18}$$

which suggests the following generalization to the non integer order μ

$$\left(\frac{d}{dx}\right)^{\mu} x^n = \frac{n!}{\Gamma(n-\mu+1)} x^{n-m} \tag{19}$$

accordingly the derivative of order ½ of $x$ writes

$$\left(\frac{d}{dx}\right)^{1/2} x = 2\sqrt{\frac{x}{\pi}} \tag{20}.$$

***ii) Derivation in closed form of different type of integrals [5]***

The method is simple and elegant and, in these introductory remarks, we will just refer to integrals we will use in the forthcoming parts of the paper.

Integrals of the type

$$I = \int_{-\infty}^{\infty} \frac{1}{\left[1+ax^2\right]^{\nu}} dx,$$
$$Re\,\nu > \frac{1}{2} \tag{21}$$

are usually treated using complex variable integration methods. It is however quite easy to obtain its explicit form in terms of Gamma functions, if one notes that the use of the Laplace Transform identity

$$\frac{1}{a^{\nu}} = \frac{1}{\Gamma(\nu)} \int_0^{\infty} e^{-sa} s^{\nu-1} ds \tag{22}$$

allows the reduction of (21) to the evaluation of a simple Gaussian integral. According to eq. (22), we can indeed cast the integral (21) in the form

$$I = \frac{1}{\Gamma(\nu)} \int_0^{\infty} ds e^{-s} s^{\nu-1} \left( \int_{-\infty}^{\infty} e^{-sax^2} dx \right) = \frac{1}{\Gamma(\nu)} \sqrt{\frac{\pi}{a}} \int_0^{\infty} e^{-s} s^{\nu-\frac{3}{2}} ds = \sqrt{\frac{\pi}{a}} \frac{\Gamma\left(\nu-\frac{1}{2}\right)}{\Gamma(\nu)} \tag{23}.$$

The same procedure can be exploited to get

$$I=\int_{-\infty}^{\infty}\frac{1}{\left[1+ax^2+bx\right]^{\nu}}dx=\sqrt{\frac{\pi}{a}}\frac{\Gamma(\nu-\frac{1}{2})}{\Gamma(\nu)}\frac{1}{\left[1+\frac{b^2}{4a}\right]^{\nu}} \tag{24}$$

$$Re\,\nu>\frac{1}{2}$$

A generalization of the Laplace transform identity (22) allows the derivation of even more complicated integrals. The identity reported below and due to Feynman [6][1]

$$\frac{1}{A^{\alpha}B^{\beta}}=\frac{1}{B(\alpha,\beta)}\int_0^1\frac{\xi^{\alpha-1}(1-\xi)^{\beta-1}}{\left[\xi A+(1-\xi)B\right]^{\alpha+\beta}}d\xi \tag{25}$$

can be exploited to reduce the integral

$$I=\int_0^{\infty}\frac{dx}{\left[1+\alpha_1x^2+\beta_1x\right]^{\nu}\left[1+\alpha_2x^2+\beta_2x\right]^{\mu}} \tag{26a}$$

to more tractable forms, which can be written on account of eqs. (25) (24) as

$$I=\frac{1}{B(\alpha,\beta)}\int_0^1 d\xi\,\xi^{\alpha-1}(1-\xi)^{\beta-1}\int_0^{\infty}\frac{dx}{\left[1+Fx^2+Gx\right]^{\alpha+\beta}}=$$

$$=\frac{\sqrt{\pi}}{B(\alpha,\beta)}\frac{\Gamma(\alpha+\beta-\frac{1}{2})}{\Gamma(\alpha+\beta)}\int_0^1 d\xi\,\xi^{\alpha-1}(1-\xi)^{\beta-1}\frac{1}{\sqrt{F}}\frac{1}{(1+\frac{G^2}{4F})^{\alpha+\beta}}, \tag{26b}$$

$$F=\alpha_1\xi+(1-\xi)\alpha_2,\quad G=\beta_1\xi+(1-\xi)\beta_2$$

---

[1] The proof is easily achieved by noting that

$$\frac{1}{AB}=\frac{1}{A-B}\int_B^A\frac{1}{z^2}dz=\int_0^1\frac{dx}{(A-B)^2\left[x-\frac{B}{A-B}\right]^2}=\int_0^1\frac{dx}{\left[xA+(1-x)B\right]^2}$$

Eq. (25) is obtained by keeping repeated derivatives with respect to $A$ and $B$ and then by assuming that the derived expression holds for any real or complex value of the order of the derivative.

The examples, given in these remarks, yield an idea of how powerful and flexible are the methods associated with the properties of the Euler functions. In the following section we will discuss how the beta function has played a key role in the evolution of the so called string theories.

## 2. EULER B-FUNCTION AND STRING THEORY

In Figure 2 we have reported the so called Chew and Frautschi (*C-F*) plot [7], proposed during the sixties of the last century. The analysis of the experimental data showed the surprising feature according to which spins of elementary particles are proportional to the square of their masses.

Albeit Figure 2 refers to meson resonances, an analogous plot holds for baryon resonances and they (meson and baryons) can be adjusted on linear trajectories, with a universal slope, whose value can be fixed around 1 $GeV^{-2}$. These trajectories are usually referred to as Regge trajectories.

An almost natural question, arising from the previous *C-F* plot behaviour, is relevant to the nature of the potential which binds the particles forming the resonances.

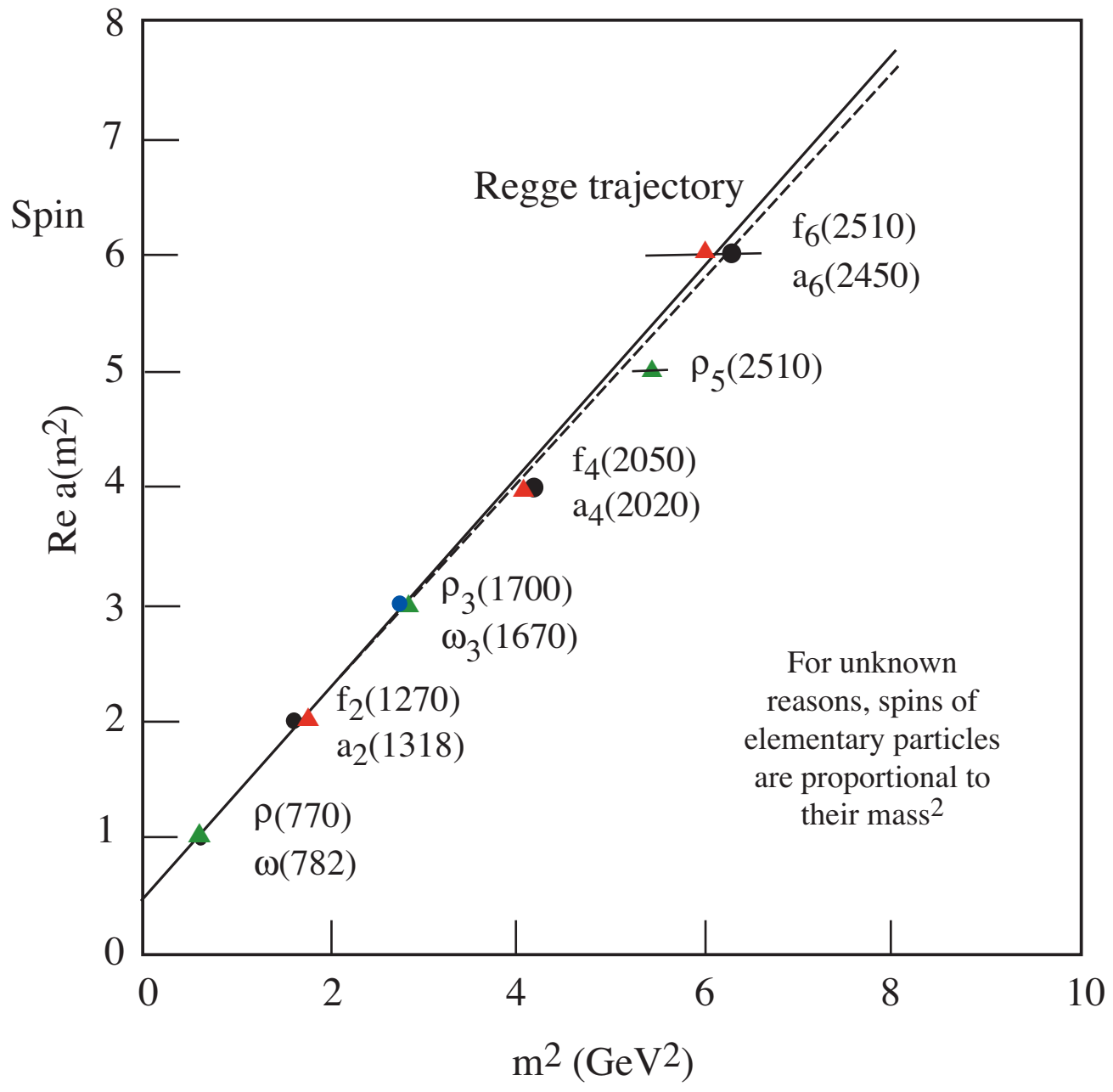

*Fig. 2 - The Chew Frautschi plot yielding the spin of mesons vs. the square of the masses*

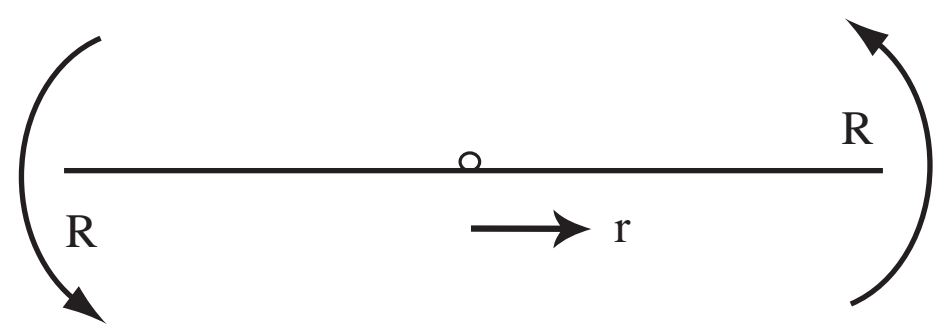

*Fig. 3 - Sketch of meson resonance composed by a rotating string*

The physical argument suggesting the form of the potential is quite straightforward. Limiting ourselves to the case of mesons, we assume that they are composed by two quarks, linked by a string (see Fig. 3). The strings nowadays called gluon flux tubes, are characterized by a linear energy density $k$, so that we can conclude that

***a) The total energy of a rotating string is***

$$E = mc^2 = 2\int_0^R \frac{k\,dr}{\sqrt{1-\left(\frac{v}{c}\right)^2}}, \frac{v(r)}{c} = \frac{r}{R} \Rightarrow$$

$$\Rightarrow 2\int_0^R \frac{k\,dr}{\sqrt{1-\left(\frac{r}{R}\right)^2}} = \pi k R \tag{27}$$

***b) The total angular momentum is***

$$J = 2\int_0^R \frac{k\,dr}{\sqrt{1-\left(\frac{v}{c}\right)^2}} r\,v\,dr = \frac{1}{2}\pi k R^2, \qquad (28).$$

Comparing a) and b) we end up with

$$J = \alpha' m^2,$$

$$\alpha' = \frac{c^4}{2\pi k} \qquad (29).$$

The string-like potential is therefore adequate to account for the phenomenology underlying the *C-F* plot, as a further remark we note that the force binding the quarks is independent of the radius and is just given by

$$\left|F_S\right| \cong \partial_R E = \frac{c^4}{2\alpha'} \qquad (30).$$

In Figure 4 we have shown an elastic scattering process, in which two incoming particles (1, 2) form two outgoing particles (3, 4). The kinematic variables, used to describe the process, are defined below, in terms of the particles four momenta

$$\begin{aligned} s &= -(p_1^{(\mu)} + p_2^{(\mu)})^2, \\ t &= -(p_1^{(\mu)} - p_4^{(\mu)})^2, \\ u &= -(p_1^{(\mu)} - p_3^{(\mu)})^2 \end{aligned} \qquad (31).$$

They are the Mandelstam variables and are not all independent since

$$s + t + u = \sum_{i=1}^{4} m_i^2 \qquad (32)$$

The variable $s$ is essentially the center of mass energy, the physical meaning of the other variables is shown in Fig. 5, where we have reported a scattering processes in the three different ***"channels"*** *s, t, u.*

The scattering amplitude can therefore be expressed in terms of these variables, which link different processes. Even though we cannot describe more precisely these concepts, we can say that, from the above description, a kind of duality emerges between interactions, described in the different channels, and that the relevant scattering amplitudes should reflect such idea.

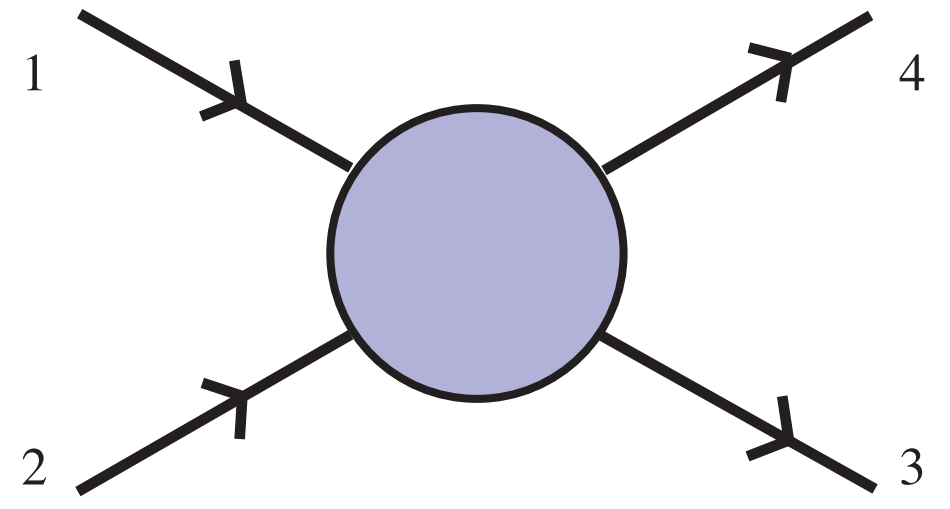

*Fig. 4 - Elastic scattering of two ingoing mesons yielding two outgoing mesons and no other particle is produced*

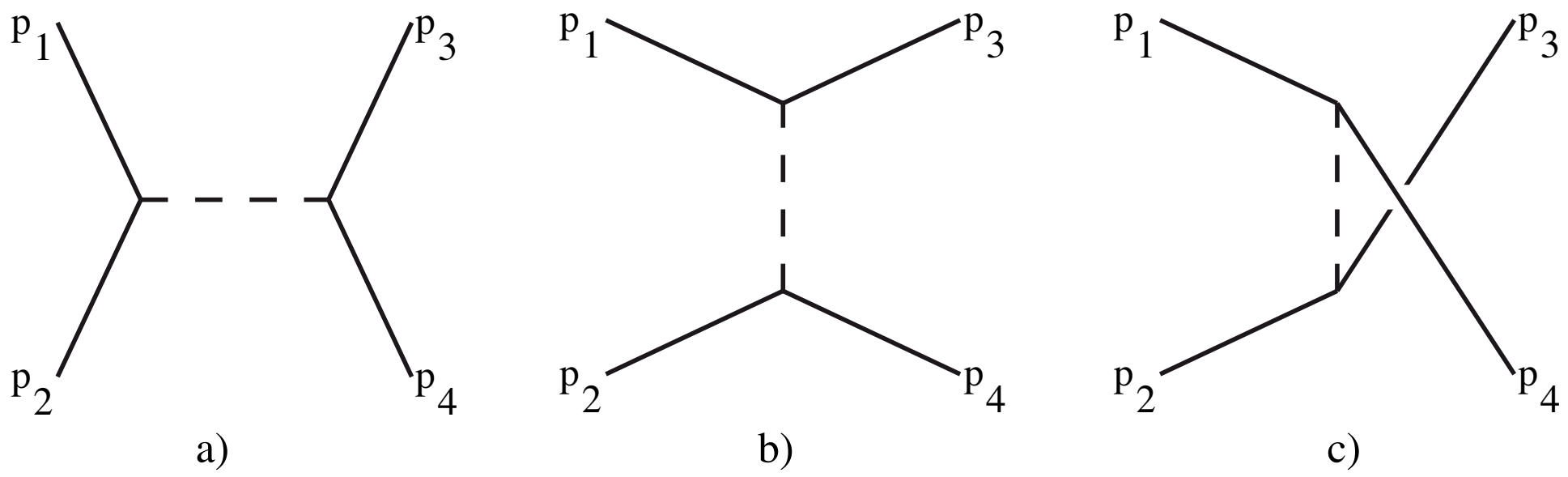

*Fig. 5- s-u-t channel scattering processes: In the s-channel the particles 1-2 couple to an unstable particle (a resonance) which eventually decays into 3, 4, in the t-channel 1 emits a particle and transform in 3, 2 absorbs the particle emitted by 1 and transforms in 4, the u-channel is the same as t with the role of 3, 4 interchanged*

The scattering amplitude depends on the trajectories $\alpha$, which can be defined in the various channels. The poles in this model play the role of resonances and the further constraint is that these poles can occur in e. g. the *s or t* channel, ***but not simultaneously***. Putting together what we have discussed so far and what we have learned in the previous section, we can conclude that a possibility of constructing such an amplitude is the use of the following definition

$$A(s,t) = \frac{\Gamma(-\alpha(s))\Gamma(-\alpha(t))}{\Gamma(-\alpha(s)-\alpha(t))} \quad (33).$$

It might be argued that the product of the $\Gamma$ functions at the denominator could be sufficient to account for the occurrence of the poles, it is therefore worth stressing that the denominator has been implanted just to avoid the occurrence of simultaneous poles, when both $\alpha(t), \alpha(s)$ are non negative integers.

The conclusion is therefore that the Euler $B$ function appears an ideally suited candidate to reproduce the scattering amplitude in terms of the Regge trajectories.

Equation (33) was first suggested by Veneziano[2] [7], and this is one of the few examples of how an intuition, relying on pure mathematical ground, had such important impact on a physical theory.

The use of the Stirling formula, discussed in the introductory section, yields e. g. [3]

---

[2] In general the scattering amplitudes writes
$T(s,t,u) = A(s,t) + A(s,u) + A(t,u)$
Which states the invariance under the interchange of the variables.

[3] In deriving eq. (34) it has been assumed that $\alpha(s) \cong \alpha_0 + \alpha' s$ and the first term has been neglected for large $s$ values.

$$lim_{\substack{s\to\infty,\\ arg(s)>0}} \quad A(s,t) \cong \Gamma(-\alpha(t)) \frac{e^{\alpha(s)}(-\alpha(s))^{-\alpha(s)-\frac{1}{2}}}{e^{\alpha(s)+\alpha(t)}(-\alpha(s))^{-\alpha(s)-\alpha(t)-\frac{1}{2}}} \cong \tag{34}$$

$$\propto \Gamma(-\alpha(t))e^{-\alpha(t)}(-\alpha' s)^{-\alpha(t)}$$

describing the amplitude behaviour in the dual *t- channel.*

It is remarkable that the Veneziano amplitude was derived as the result of an "educated" guess, based on general physical requirements fulfilled by a well known special function. The guess was later proved correct by Nambu and co-workers [7], who derived it from first principles, based on string potentials.

It is even more amazing that a mathematical tool conceived in the XVII century was the opening key for the String theory, which, in the words of Daniele Amati, provides

***…A theory of 21-th 3century fallen by chance in the 20-th century.***

## 3. THE EULER-RIEMANN FUNCTION AND THE CASIMIR EFFECT

In the following we will discuss the methods employed by Euler to treat converging and diverging series.

The first example is the sum of the series [8]

$$\begin{aligned} S_1 &= 1-1+1+\ldots = \frac{1}{2} \\ S_2 &= 1-2+3-4+\ldots = \frac{1}{4} \end{aligned} \tag{35}.$$

The proof of the "validity" of the previous idetities is sketched below

$$\begin{aligned} &S(\alpha,t) = \sum_{s=0}^{\infty}(-\alpha)^n t^n = \frac{1}{1+\alpha t} \Rightarrow S(1,1) = S_1 = \frac{1}{2} \\ &\partial_t S(1,t) = \sum_{n=1}^{\infty}(-1)^n n\, t^{n-1} = -\frac{1}{(1+t)^2}\,|_{t=1} \Rightarrow -\partial_t S(1,t)\,|_{t=1} = S_2 = 1-2+3-4+\ldots = \frac{1}{4} \end{aligned} \tag{36}$$

Our "proof" is clearly flawed by the fact that we have disregarded th that convergence, in classical sense, is ensured for $|t| < 1$ only.

Notwithstanding if we continue to trust the second of eqs. (35), we can derive even more disturbing identities.

By denoting with $O$ and $E$ the infinite sums of even and odd positive integers, we find

$$\begin{aligned}
&\Sigma = O + E \\
&O = 1 + 3 + 5 + ... \\
&E = 2 + 4 + 6 + ... = 2\Sigma \\
&S_2 = O - E = \\
&\Rightarrow 2E = \Sigma - S_2 \Rightarrow \\
&4\Sigma = \Sigma - S_2 \\
&\Rightarrow \Sigma = -\frac{S_2}{3} \\
&\Rightarrow 1 + 2 + 3 + ... = -\frac{1}{12}
\end{aligned} \tag{37}$$

Even though we have obtained the previous results as the consequence of dirty manipulations, they have a grain of truth as it will be shown in the following.

To this aim we introduce the Euler-Riemann (E-R) function, defined by the infinite series

$$\begin{aligned}
&\zeta(x) = \sum_{n=1}^{\infty} \frac{1}{n^x}, \\
&Re(x) > 1
\end{aligned} \tag{38).}$$

It is evident that if we relax the condition $Re\, x > 1$ and accept the validity of eqs. 35-37 then $\zeta(-1) = -1/12$.

The use of the Laplace transform identity

$$\frac{1}{n^x} = \frac{1}{\Gamma(x)} \int_0^\infty e^{-n s} s^{x-1} ds \tag{39}$$

allows the derivation of the E-R function integral representation

$$\begin{aligned}
&\zeta(x) = \frac{1}{\Gamma(x)} \int_0^\infty \sum_{n=1}^{\infty} e^{-n s} s^{x-1} ds = \frac{1}{\Gamma(x)} \int_0^\infty \frac{e^s}{e^s - 1} s^{x-1} ds, \\
&Re(x) > 1
\end{aligned} \tag{40}$$

Its behaviour, for positive real $x$-values, is reported in Fig. 6.

The E-R function plays a central role in a variety of physical problems, a very well known example is the black body radiation formula [9] (see Fig. 7), and the associated energy per unit volume, which, as easily verified, can be written as

$$u(\nu,T)=\frac{8\pi h\nu^3}{c^3}\frac{1}{e^{\frac{h\nu}{KT}}-1},$$
$$\int_0^\infty u(\lambda,T)\,d\lambda=\frac{U}{L^3}\Rightarrow\int_0^\infty\frac{x^3}{e^x-1}=\zeta(4)\Gamma(4) \tag{41}$$

The values of the E-R function, for even values of the argument, can be evaluated using the properties of the Bernoulli numbers. To this aim we remind that the Bernoulli numbers are defined through the generating function

$$\frac{t}{e^t-1}=1+\sum_{k=2}^{\infty}\frac{t^k}{k!}B_k,$$
$$B_1=-\frac{1}{2} \tag{42a}$$

and that they are explicitly defined as

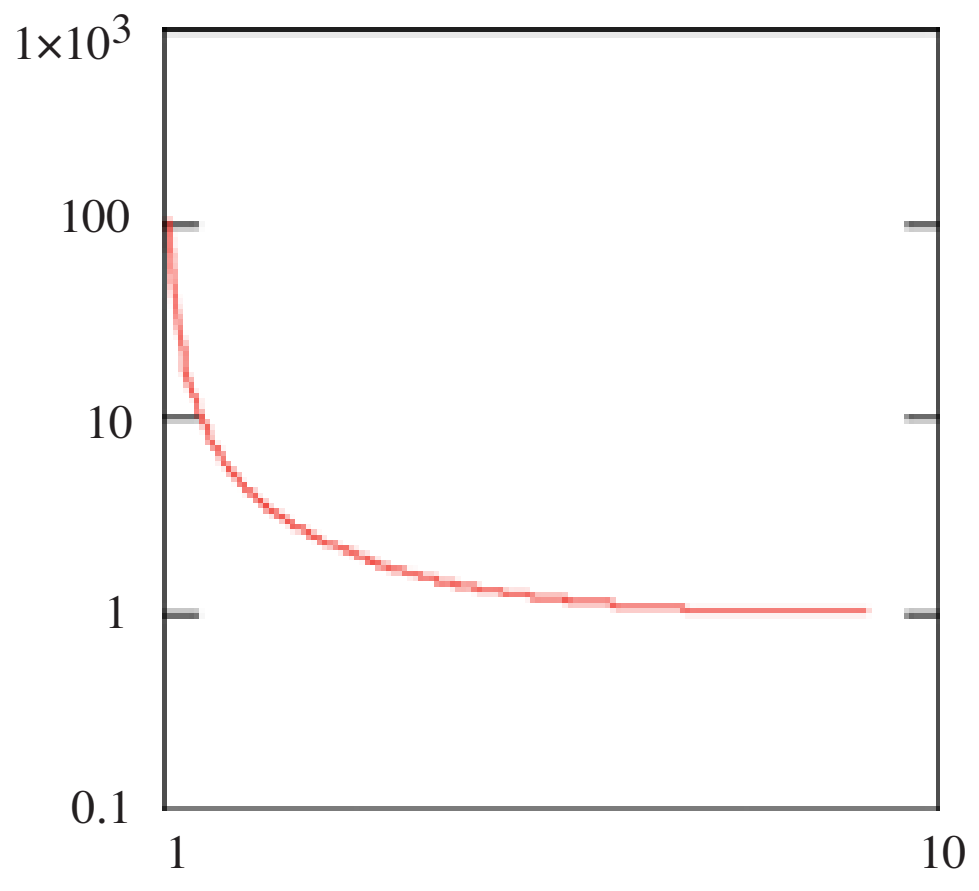

*Fig. 6 - Log-plot of $\zeta(x)$*

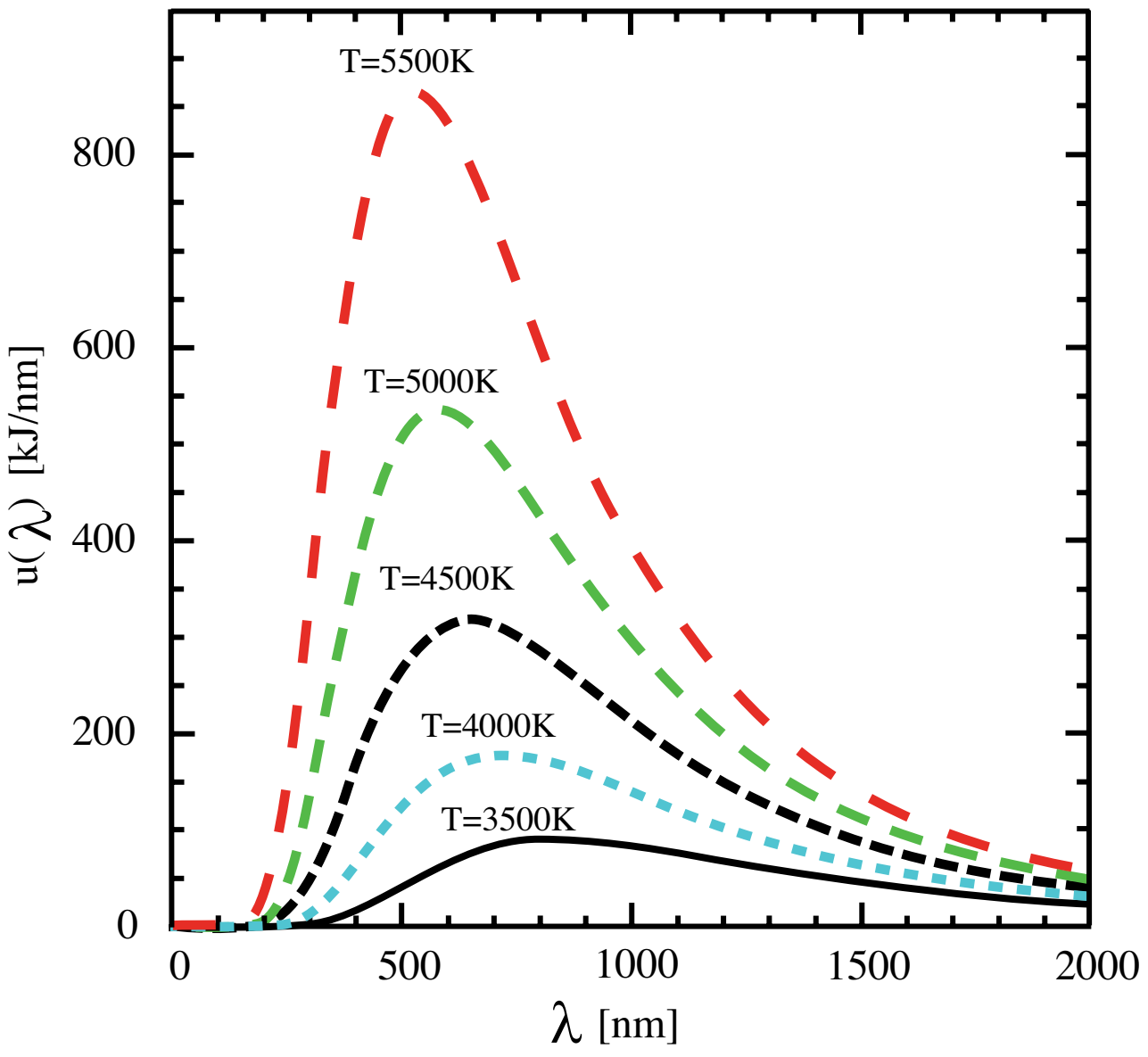

*Fig. 7 - Black body radiation curve*

$$B_k = \frac{(-1)^k k}{2^k - 1} \sum_{i=1}^{k} 2^{-i} \sum_{j=0}^{i-1} (-1)^j \binom{i-1}{j} (j+1)^{k-1},$$

$$k \geq 1, \qquad (42b).$$

$$B_0 = 1$$

According to eq. (42a) the following expansion is easily proved

$$ctg(z) = \frac{1}{z} + \sum_{k=1}^{\infty} \frac{(-1)^k 2^{2k} B_{2k}}{(2k)!} z^{2k-1} \qquad (43a)$$

Therefore, using the further identity

$$ctg(z) - \frac{1}{z} = \sum_{n=1}^{\infty} \frac{2z}{z^2 - (n\pi)^2} = -2 \sum_{k=0}^{\infty} \frac{z^{2k-1}}{\pi^{2k}} \zeta(2k) \qquad (43b)$$

we find, after comparing the like power $z$-coefficients,

$$\zeta(2m) = \frac{(-1)^{m-1} 2^{2m-1} B_{2m}}{(2m)!} \pi^{2m} \qquad (44).$$

The E-R function can be analytically continued to values of the variable such that $\mathrm{Re}(x) \leq 1$ (see Fig. 8). The use of the formula (due to Riemann)

$$\zeta(1-s) = 2^{1-s} \pi^{-s} \Gamma(s) cos\left(\frac{s\pi}{2}\right) \zeta(s), \qquad (45)$$

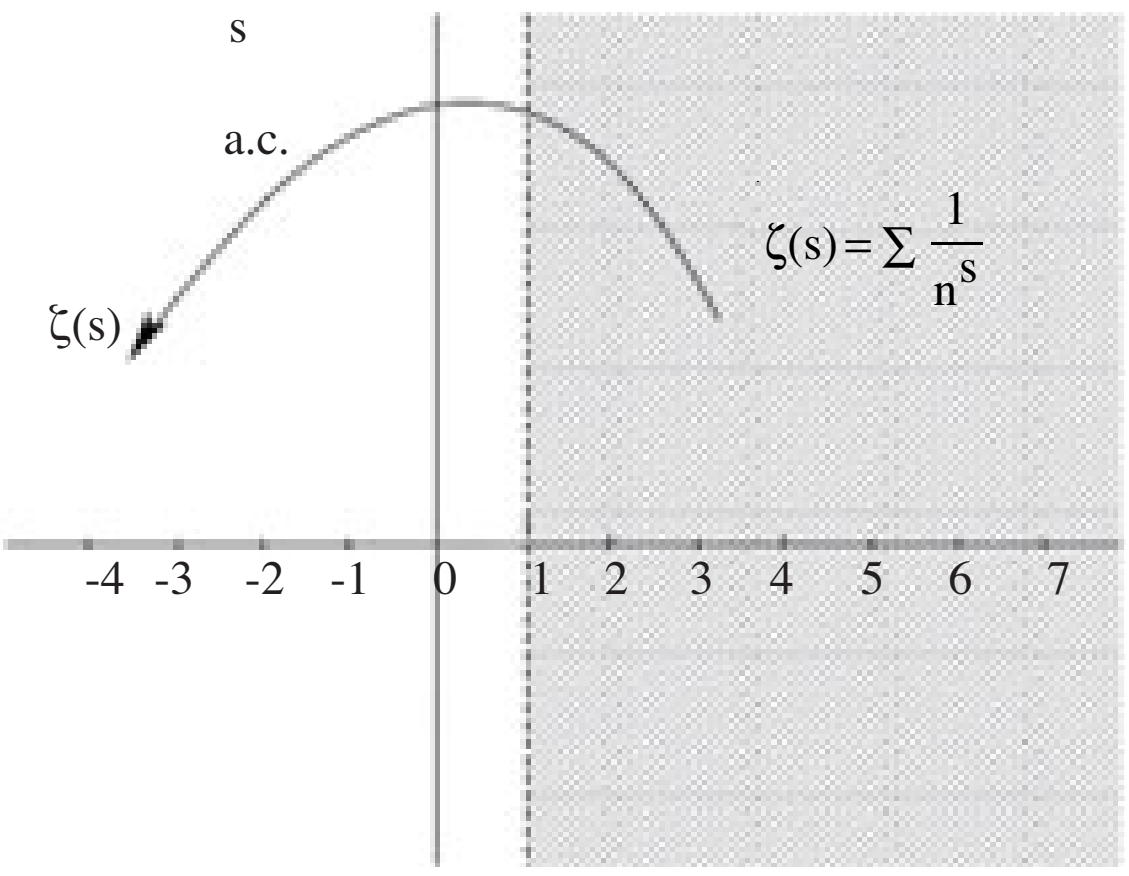

*Fig. 8 - Analytical continuation of the Riemann function*

yields such an analytical continuation and provides a meaning for the divergent series, obtained at the beginning of this section, we find indeed

$$s = 2 \Rightarrow$$

$$\Rightarrow \zeta(-1) = \sum_{n=1}^{\infty} \frac{1}{n^{-1}} = \sum_{n=1}^{\infty} n = \tag{46}$$

$$-\frac{1}{2\pi^2}\Gamma(2)\zeta(2) = -\frac{1}{2\pi^2}\sum_{n=1}^{\infty}\frac{1}{n^2} = -\frac{1}{12}$$

This, apparently foolish, conclusion has consequences of paramount importance in Physics, in particular in the theory of the renormalization and allows the possibility of providing a meaning for divergent integrals in quantum electrodynamics, like those emerging in the study of the Casimir effect, which is a manifestation of forces of genuine quantum nature [10].

In Figure 9 we have reported two conducting plates, separated by a distance *a*, for reasons of mode confinement, not all the modes of the vacuum electromagnetic field will be contained within the conducting plates, those with longer wave lengths will be "confined" outside. The energy of the field is larger than that inside, the plates are therefore experiencing an attractive force, due to the electromagnetic pressure of radiation.

The force pressure, acting on the faces, is just given by the derivative (with respect to the coordinate *a*) of the energy density inside, namely

$$\bar{F}_c = -\partial_a \frac{\langle E \rangle}{A} \tag{47}$$

Where $\langle E \rangle / A$ is the average energy density (*A* is the plate's surface). Without entering into the details of the calculations, we note that it is provided by

$$\frac{\langle E \rangle}{A} = -\frac{\hbar c \pi^2}{6a^3} \sum_{n=1}^{\infty} n^3 \propto \zeta(-3),$$
$$\zeta(-3) = \frac{1}{8\pi^4}\Gamma(4)\zeta(4) \qquad (48).$$

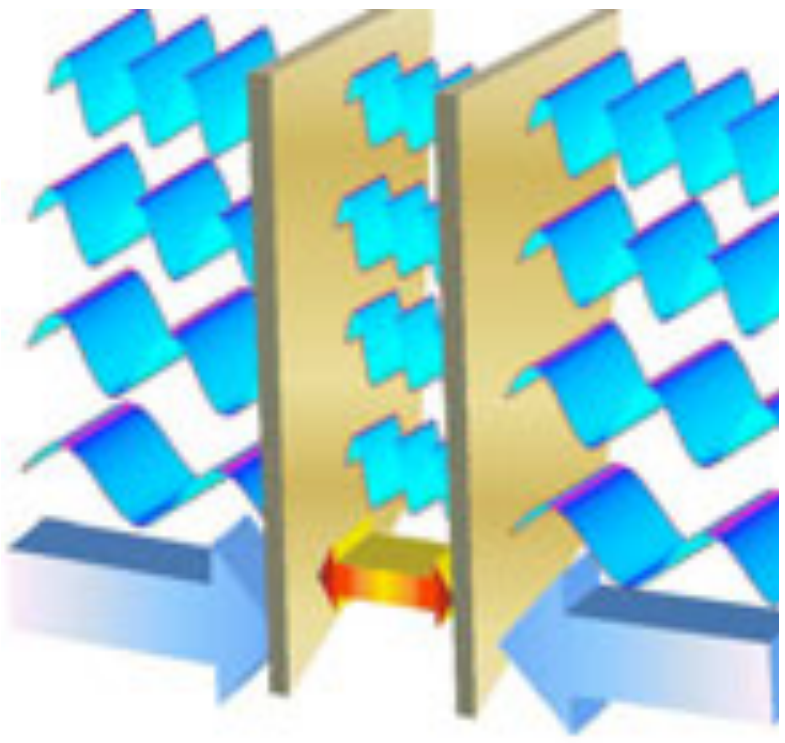

*Fig. 9 - Conducting plates and Casimir effect*

It happens therefore that we *"regularize"* infinite sums using a mathematical procedure derived within the context of the analytical continuation of the E-R function.

## 4. THE EULER-RIEMANN FUNCTION AND THE THEORY OF RENORMALIZATION

The theory of the renormalization can be defined as the this definition is move appropriate for the underlying mathematical technicalities known as regolarization

***art of removing the infinities***

To better clarify what we mean by "removing infinities", we consider an example from elementary calculus. It is well known that

$$I_n = \int x^n dx = \frac{1}{n+1} x^{n+1} + c \qquad (49)$$

Which diverges if we take the limit $n \to -1$. We can obtain a finite result if we "subtract" the infinite by properly choosing the constant *c.* By setting indeed[4]

$$I_n = \frac{1}{n+1} x^{n+1} - \frac{1}{n+1} \qquad (50)$$

we obtain (see Fig. 10)

$$lim_{n \to -1} I_n = ln(x) \qquad (51)$$

as it must be, since $\int \frac{1}{x} dx = \ln(x)$.

4 This argument was developed during an informal discussion of one of the Authors (G. D. ) with T. E. O. Hermsen.

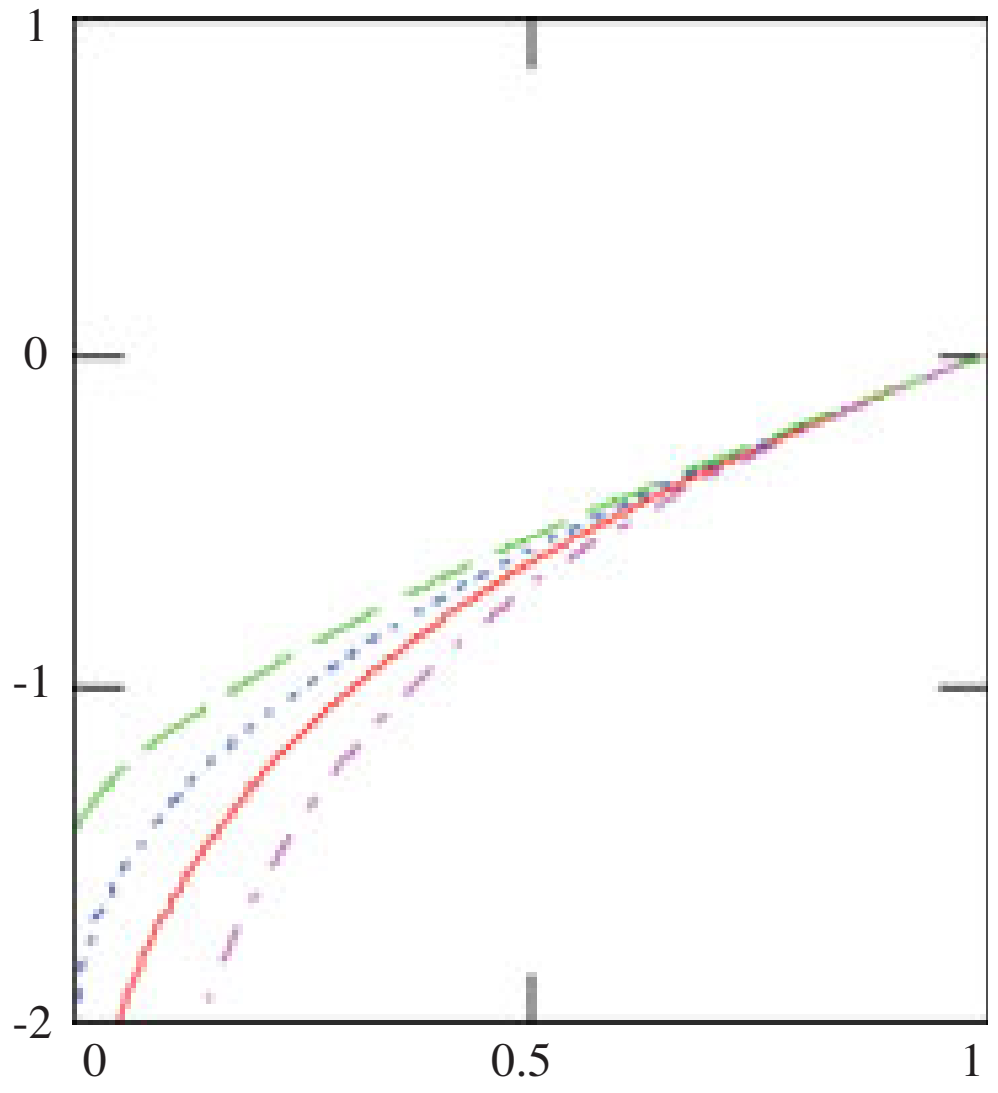

*Fig. 10* - In vs. x *for different values of n (=-0.3 dash, =-0.5 dot, =- 0.7 continuous =-1 dash-dot)*

Let us now consider the following infinite sum

$$\Sigma(x) = \sum_{n=0}^{\infty} f(x+n) \tag{52}$$

which, on account of the identity $e^{\lambda \partial_x} f(x) = f(x+\lambda)$, can be written as

$$\Sigma(x) = \sum_{n=0}^{\infty} e^{n \partial_x} f(x) = -\frac{\partial_x}{e^{\partial_x} - 1} F(x),\ f(x) = \partial_x F(x) \tag{53}.$$

The use of the Bernoulli numbers generating function (42a) yields

$$\Sigma(x) = -\sum_{n=0}^{\infty} \frac{B_n}{n!} \partial_x^n F(x) \tag{54}$$

If for example $f(x) = x^m$, we obtain that $\Sigma(0) = -B_{m+1}/m+1$, which is a finite result and coincides with $\zeta(-m)$. Also in this case we have subtracted from an expression, giving rise to a divergent term, an other divergence of the same "order".

In more crude terms we state that we have managed to reduce our problem to an operation of the type

$$\Re = (\infty + a) - \infty$$

Talking of dirty tricks!!!

In an attempt of providing a more rigorous formulation for the subtraction of infinities, we define the following Ramanujan, renormalized sum of a diverging series as

$$\sum_n^{\Re} a(n) = lim_{n\to\infty}\left(\sum_{n=1}^{N} a(n) - \int_1^N a(t)\,dt\right) \tag{55}$$

The above formula allows to assign the value $\gamma$ (the Euler-Mascheroni constant) to the diverging series $\sum_{n=1}^{\infty}\frac{1}{n}$ while for divergent oscillating series we find

$$\sum_n^{\Re} sin(n\,t) = \frac{1}{2}cot\left(\frac{t}{2}\right) - \frac{cos(t)}{t} \tag{56}$$

The moral, beyond these examples is that we have "regularized" an integral or a divergent sum by subtracting a term which has the "same degree" of infinity of the diverging term.

To be honest and avoid any confusion, we must stress again that, even though eq. (54) represents a term by term cancellation, the perplexities around the procedure remain, we have eliminated an infinite by subtracting to it an other infinite, mathematicians may be horrified[5] but this is, in a nut shell, the bare essence of the renormalization.

Infinities are often encountered in quantum (as well as in classical) Physics, one example is provided by the Feynman diagram reported in Fig. 11, regarding the corrections to the scattering $e^+e^- \to \mu^+\mu^-$ [7].

The momentum conservation requires the summation over all the momenta around the loop, thus yielding a contribution proportional to ($\hbar = c = 1$)

$$I_{loop} \propto \int \frac{d^4k}{(2\pi)^4}\frac{1}{\left[k^2 - m^2\right]\left[(k-p)^2 - m^2\right]} \tag{57}$$

This is clearly diverging, neglecting the mass $m$ and momentum $p$ we find

$$I_{loop} \propto \int \frac{d^4k}{(2\pi)^4}\frac{1}{k^4} \cong \int_{k_{min}}^{\infty}\frac{dk}{k} = ln(\infty) - ln(k_{min}) \tag{58}$$

and the above divergence is usually referred as ***Ultra-Violet singularity***.

---

5 We will come back to the criticisms in the concluding part of the paper.

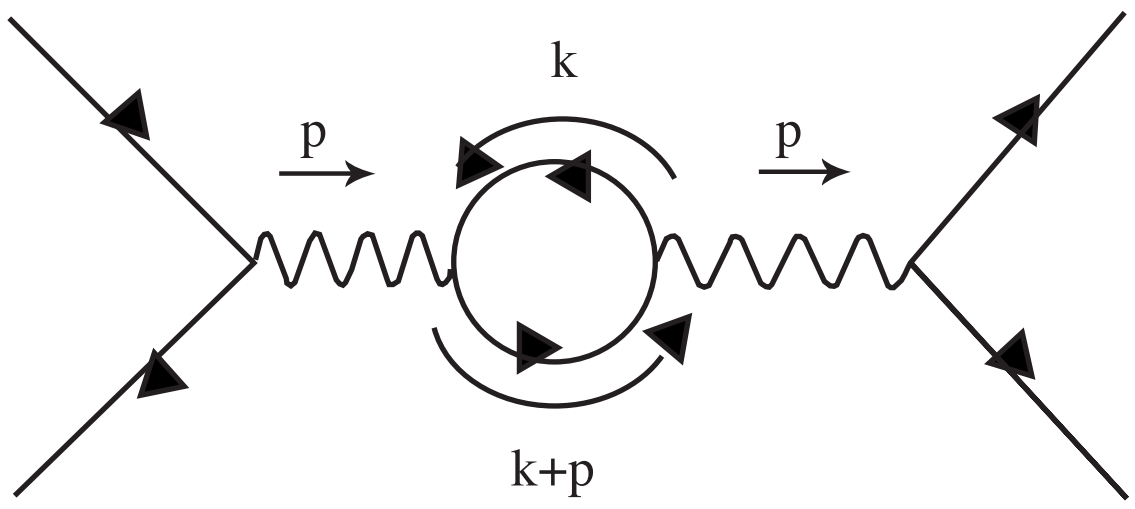

*Fig. 11 - Feynmann diagram for the calculation of the cross section of the* $e^+e^- \to \mu^+\mu^-$ *process*

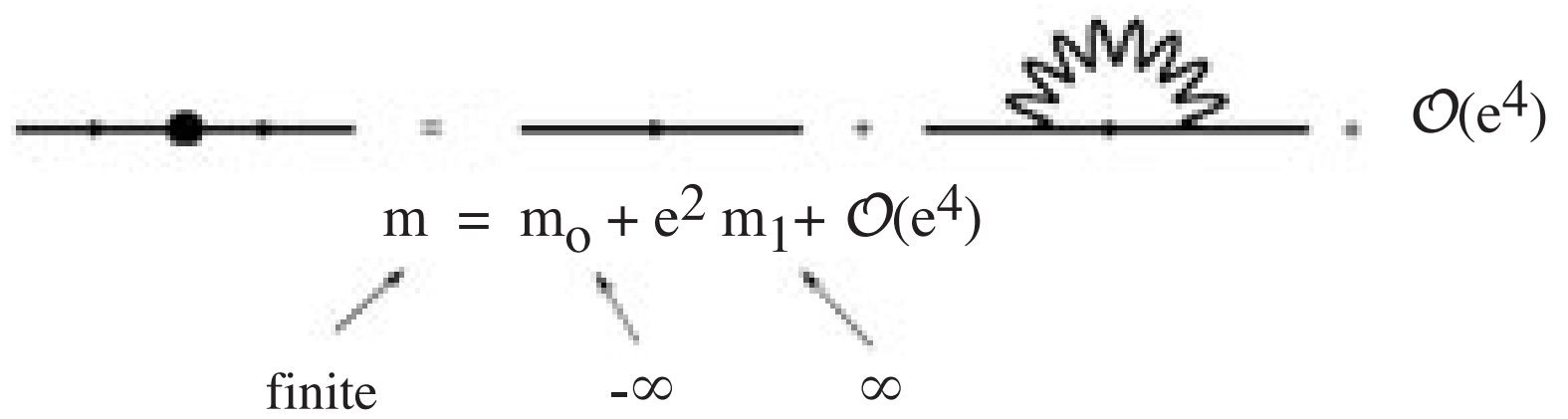

*Fig. 12 - One loop correction of the electron mass*

What makes a theory useful, namely capable of making reliable predictions, is just the fact that it contains the possibility of eliminating the infinities. Why this is necessary is expressed by the example illustrated by Fig. 12, where we report the contribution at the lowest order corrections to the bare electron mass. Since the loop yields a divergence, the bare mass too should be divergent, in order to avoid the infinite appearing in the corrections.

We have already discussed a fairly simple mathematical examples of renormalization regarding elementary functions and we want to stress that the paradox leading to finite sums of the type $\zeta(-s)$ is simply due to the fact that actually $\zeta(-s) \neq \sum_{s=1}^{\infty} \frac{1}{n^{-s}}$ but it represent its ***"renormalized"*** counterpart.

In general the integrals which should be cured in QED are of the type

$$I(m.\Lambda) = \int_0^{\Lambda} p^m dp \tag{59}$$

The parameter $\Lambda$ appearing in eq. (59) is called the regulator and is an essential element in the theory of renormalization The integral (59) is diverging for $\Lambda \to \infty$, therefore much care should be devoted in dealing with these integrals and then take the limit to eliminate the divergences.

The quantity $p$ is in general linked to the momentum of the exchanged particle, if it is a photon $p \propto 1/\lambda$ and, depending on the sign of *m,* we may have infrared (low momentum) or ultra violet (large momentum) divergences.

The first who suggested a way out to the problem of infinities by the use of the E-R function was Schwinger [11], who noted that, if we transform the above integral into a sum, we find

$$lim_{\Lambda\to\infty} I(m,\Lambda) \cong \zeta(-m) \qquad (60)$$

Which is finite for any positive $m$. The approximation is too crude to get always consistent results and cannot be extended to any real (positive or negative) value of *m.*

The use of the Trapezoid integral method allows a more rigorous handling of the integrals of eq. (59). We find indeed [7]

$$I(m,\Lambda) = \int_0^\Lambda p^m dp \cong \frac{\Lambda^m}{2} + \sum_{n=1}^{\Lambda} n^m - \sum_{r=1}^{\infty} \frac{B_{2r}}{(2r)!} a_{m,r} \Lambda^{m-2r-1},$$

$$a_{m,r} = \frac{\Gamma(m+1)}{\Gamma(m-2r+2)}, \qquad (61).$$

By noting that

$$\Lambda^s = s\int_0^\Lambda p^{s-1} dp = s I(s-1,\Lambda) \qquad (62)$$

We can cast eq. (61) in he form of the following recurrence

$$I(m,\Lambda) \cong \frac{m}{2} I(m-1,\Lambda) + \zeta(-m)$$
$$- \sum_{r=1}^{\infty} \frac{B_{2r}}{(2r)!} a_{m,r} (m-2r+1) I(m-2r,\Lambda) \qquad (63)$$

which is a refinement of the original Schwinger's suggestion, as proposed more recently by Elizalde, Garcia and Hartle [11]. In this way all the integrals of the type $I(m,\Lambda)$ can be written as linear combination of $\zeta(-2r-1)$. From eq. (62) we get indeed the recusions

$$I(0,\Lambda) \cong \zeta(0) = -\frac{1}{2},$$

$$I(1,\Lambda) \cong \frac{1}{3} I(0,\Lambda) + \zeta(-1),$$

$$I(3,\Lambda) = \frac{1}{2}(I(0,\Lambda) - a_{2,1}B_2) + \zeta(-1), \tag{64}$$

...

We consider now the case of negative $m$ values, we introduce a second regulator $\varepsilon$ and write (see eqs. (21-24))

$$I(m,\Lambda,\varepsilon) = \int_0^{\Lambda} \frac{1}{(\varepsilon + p)^m} dp = \frac{1}{\Gamma(m)} \int_0^{\Lambda} dp \int_0^{\infty} e^{-s(\varepsilon+p)} s^{m-1} ds =$$
$$= \frac{\varepsilon^{-(m-1)}}{\Gamma(m)} \int_0^{\infty} \sigma^{m-1} e^{-\sigma} d\sigma \left[ \int_0^{\Lambda \varepsilon^{-1}} e^{\sigma\,\omega}\, d\omega \right] \tag{65}$$

By expanding the integral in square brackets, we eventually end up with

$$I(m,\Lambda,\varepsilon) \cong \frac{\varepsilon^{-(m-1)}}{\Gamma(m)} \sum_{n=0}^{\infty} \frac{\Gamma(m+n)}{n!} I\left(n, \frac{\Lambda}{\varepsilon}\right) \tag{66}$$

which again yields the integral in terms of E-R functions. One can clearly rise doubts on the convergence of the series on the left of eq. (66). This is a problem indeed and the correctness of the procedure should always be checked a posteriori.

## 5. THE EULER GAMMA FUNCTION AND RENORMALIZATION TECHNIQUES

As we have already mentioned that Feynman diagrams involving loop corrections, like that reported in Fig. 13 [7], requires the evaluation of integrals of the type

$$I(d,n) = \int \frac{d^d k}{\left[k^2 + 1\right]^n} \tag{67}$$

where $d$ denotes the dimensionality of the space, where we are performing the integration.

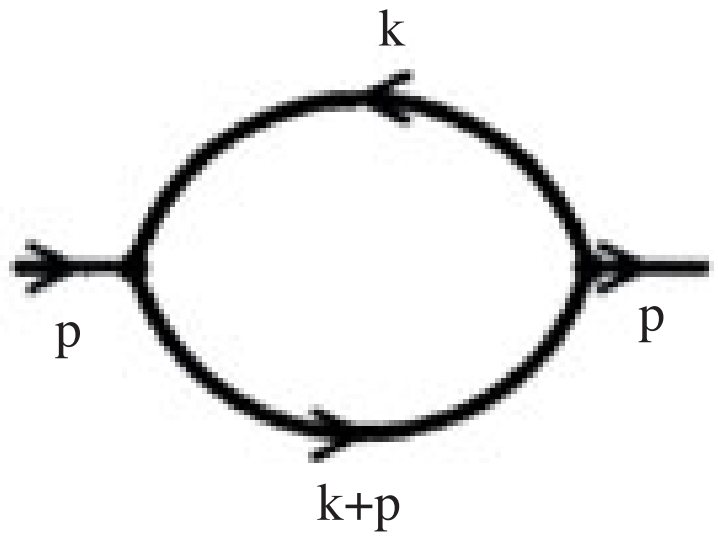

*Fig. 13 - Feynman diagram for one loop corrections*

The integral (67), even though apparently complicated, can be treated using the methods discussed in the introductory part of this paper. The Laplace transform technique allows indeed the cast eq. (67) in the form

$$\begin{aligned} I(d,n) &= \frac{1}{\Gamma(n)} \int d^d k \int_0^\infty e^{-s} e^{-sk^2} s^{n-1} ds = \\ &= \frac{1}{\Gamma(n)} \int_0^\infty e^{-s} s^{n-1} ds \left[ \int d^d k\, e^{-sk^2} \right] \end{aligned} \tag{68}$$

The integral in the square brackets, over the *d*-dimensional volume, is easily worked out as

$$\int d^d k\, e^{-sk^2} = \left[ \int_{-\infty}^{\infty} dk_x e^{-sk_x^2} \right]^d = \left( \frac{\pi}{s} \right)^{\frac{d}{2}} \tag{69}$$

We can finally express $I(d,n)$ in terms of the gamma functions as follows

$$I(d,n) = \frac{\pi^{\frac{d}{2}}}{\Gamma(n)} \Gamma(n - \frac{d}{2}) = \frac{\pi^{\frac{d}{2}}}{\Gamma\left(\frac{d}{2}\right)} B\left(n - \frac{d}{2}, \frac{d}{2}\right) \tag{70}$$

The integral in eq. (69) converges only if $d<2n$, we note however that, by setting $\varepsilon = 1 - d/2$, we can write

$$I(d,1) = 4\pi^{1-\varepsilon} \varepsilon \Gamma(\varepsilon) \tag{71}$$

The other integrals for $n > 1$ are all proportional to eq. (71), as easily verified, we find e. g. for $n=2$

$$I(d,2) = \varepsilon I(d,1) \tag{72}$$

and, for larger $n$ values, we can establish straightforward recurrences.

The conclusion is that we have individuated the source of the divergences, which manifest themselves as pole in the complex plane and can be therefore subtracted by direct methods.

As a further example, we consider the first order loop integral appearing in $\varphi^4$ perturbative theories, ($\lambda$ is a coupling constant)

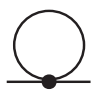

$$= -\lambda \int \frac{d^D p}{(2\pi)^D} \frac{1}{p^2 + m^2} \tag{73}$$

It can be written in terms of *I(d,1)*, and the relevant divergences located at positive even $d$ values, can be subtracted as previously illustrated.

Second loop integrals

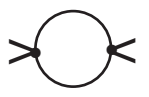

$$= -\lambda^2 \int \frac{d^D p}{(2\pi)^D} \frac{1}{p^2 + m^2} \frac{1}{(p+k)^2 + m^2} \tag{74}$$

can be evaluated by means of analogous tricks, involving the properties of the gamma function too, they can be indeed reduced to the simplest form

$$\frac{\lambda^2}{(4\pi)^{\frac{d}{2}}} \frac{\Gamma(2-\frac{d}{2})}{\Gamma(2)} \int_0^1 dx \frac{1}{\left[k^2 x(1-x) + m^2\right]^{2-\frac{d}{2}}} \tag{75}$$

after using the already quoted identity (see eq. (25) and footnote 1)

$$(AB)^{-1} = \int_0^1 \frac{dx}{\left[xA + (1-x)B\right]} \quad (76).$$

In this section we have seen how different methods of operational nature can be combined and applied to Physics problem in the following sections we will add further examples.

## 6. MATHEMATICS OR MATHEMAGICS

Some Mathematicians (and even Physicists) including Euler, Ramanujan and Feynman are some times called Mathemagicians, for their attitude in giving a meaning to absurd.

Symbols in Mathematics are like the words of the ordinary language. These symbols are usually entangled to form a phrase with a definite meaning. Words, if put together without a proper organization, looses any meaning, this happens for mathematical symbols too.

Meaning is, however, not absolute but relative, since in many cases it needs a key of interpretation, in some cases the absence of meaning may be only apparent.

Mathemagicians have the ability of interpreting the key of apparently disconnected symbols, thus yielding a meaning to seemingly foolish expressions.

Just to give an example we consider the following relation [12]

$$\begin{aligned} & F(x) = \hat{E}_{\partial_x} f(x), \\ & \hat{E}_{\partial_x} = \left[\partial_x\right]! \end{aligned} \tag{77}$$

Which is, at first sight, a meaningless operation, since we are not acquainted with the definition of an operator, which is the factorial of the derivative operator.

We remind however that the use of the gamma function has provided a meaning for the extension of the factorial operation to any real or complex number and to the definition of the derivatives of fractional orders. This last extension has required two distinct logical steps

**1) recognize the fact that the derivative can be treated as a common algebraic quantity**

**2) the use of the properties of the factorial for any real number**

The same point of view can be adopted to define the operator appearing in eq. (77), we can therefore using the gamma function formally write

$$\left(\partial_x\right)! = \Gamma(\partial_x + 1) = \int_0^\infty e^{-\xi} \xi^{\partial_x} d\xi \tag{78}$$

and, by setting $\xi^{\partial_x} = e^{ln(\xi)\partial_x}$ , we find

$$\hat{E}_{\partial_x} f(x) = \int_0^\infty e^{-\xi} f(x + \ln(\xi))\, d\xi \qquad (79),$$

and conclude that the action of the factorial derivative on a given function can be expressed in terms of the logarithmic convolution reported in eq. (79), the meaning of such operation is therefore well defined, even though apparently crazy.

Other identities, disturbing as the Munk scream (Fig. 14), may be reconciled with a non eicsoteric interpretation, as disussed below.

In the previous sections we have seen that the regularization of divergent series is a consequence of analytic continuation techniques, but Euler obtained the same result without any recourse to methods of complex analysis, which were largely unknown to him. Just repeating the previously outlined procedure, we can adopt the Euler method to evaluate sums of the type [13]

$$\sum_{n=1}^{\infty}(-1)^n n^k = \partial_\xi^k \sum_{n=1}^{\infty}(-1)^n e^{-n\xi} \,|_{\xi=0} = -\partial_\xi^k \frac{1}{e^\xi + 1} \,|_{\xi=0} \qquad (80)$$

and by noting that (see eq. (42 b))

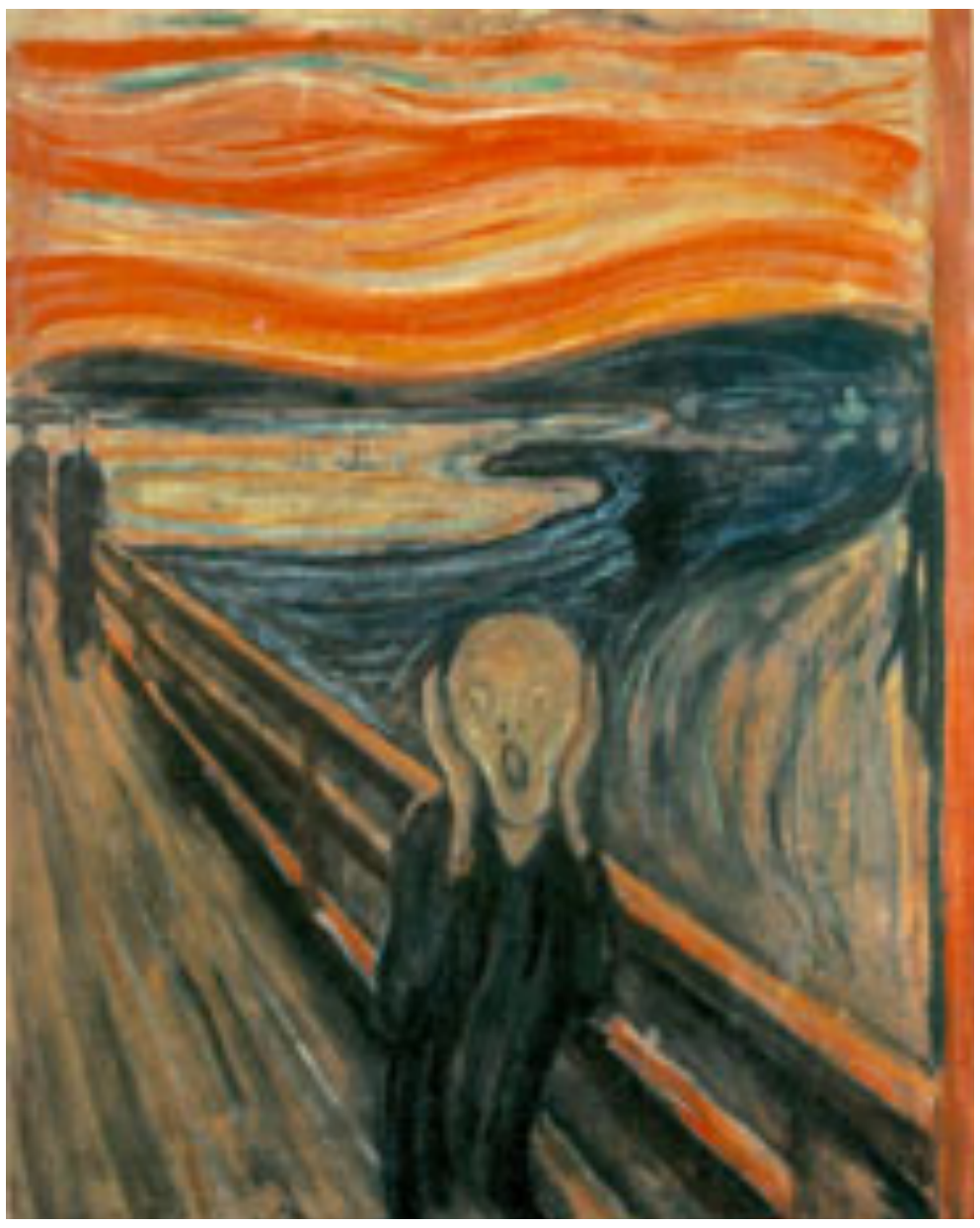

*Fig. 14 - The Munk Scream (According to the Authors' interpretation it can be understood as the reaction of a rigorist to identities of the type (77) and (81))*

$$\frac{1}{e^{\xi}+1}=\frac{1}{e^{\xi}-1}-\frac{2}{e^{2\xi}-1}=\sum_{k=1}^{\infty}\frac{B_{k+1}}{(k+1)!}\xi^{k}+\frac{1}{\xi} \qquad (81a).$$

we can conclude that

$$\sum_{n=1}^{\infty}(-1)^{n}n^{k}=\frac{(1-2^{k+1})}{k+1}B_{k+1} \qquad (81b)$$

The use of the techniques discussed in the paper yields other identities as e.g.

$$\infty!=\sqrt{2\pi} \qquad (82)$$

whose proof goes as sketched below

$$lim_{n\to\infty}(n!)=lim_{n\to\infty}e^{\sum_{s=1}^{n}ln(s)} \qquad (83a)$$

being

$$\sum_{s=1}^{\infty}ln(s)=-\zeta'(0)=ln(\sqrt{2\pi}) \qquad (83b)$$

we obtain the identity (82) .

Regarding eq. (83b) we note that it just follows from a renormalization procedure of the type

$$\begin{aligned}\sum\nolimits_{n}^{\Re}ln(n)&=lim_{n\to\infty}\left[\sum_{n=1}^{N}ln(n)-\int_{1}^{N}ln(t)\,dt)\right]\\&=lim_{n\to\infty}\left[ln(N!)-(N\,ln(N)-N)\right]\end{aligned} \qquad (83c)$$

whose graphical interpretation is reported in Fig. 15.

We can exploit the concept of factorial derivative to get a meaning for the “wildly” divergent series

$$\Delta=\sum_{n=1}^{\infty}(-1)^{n}n! \qquad (84a),$$

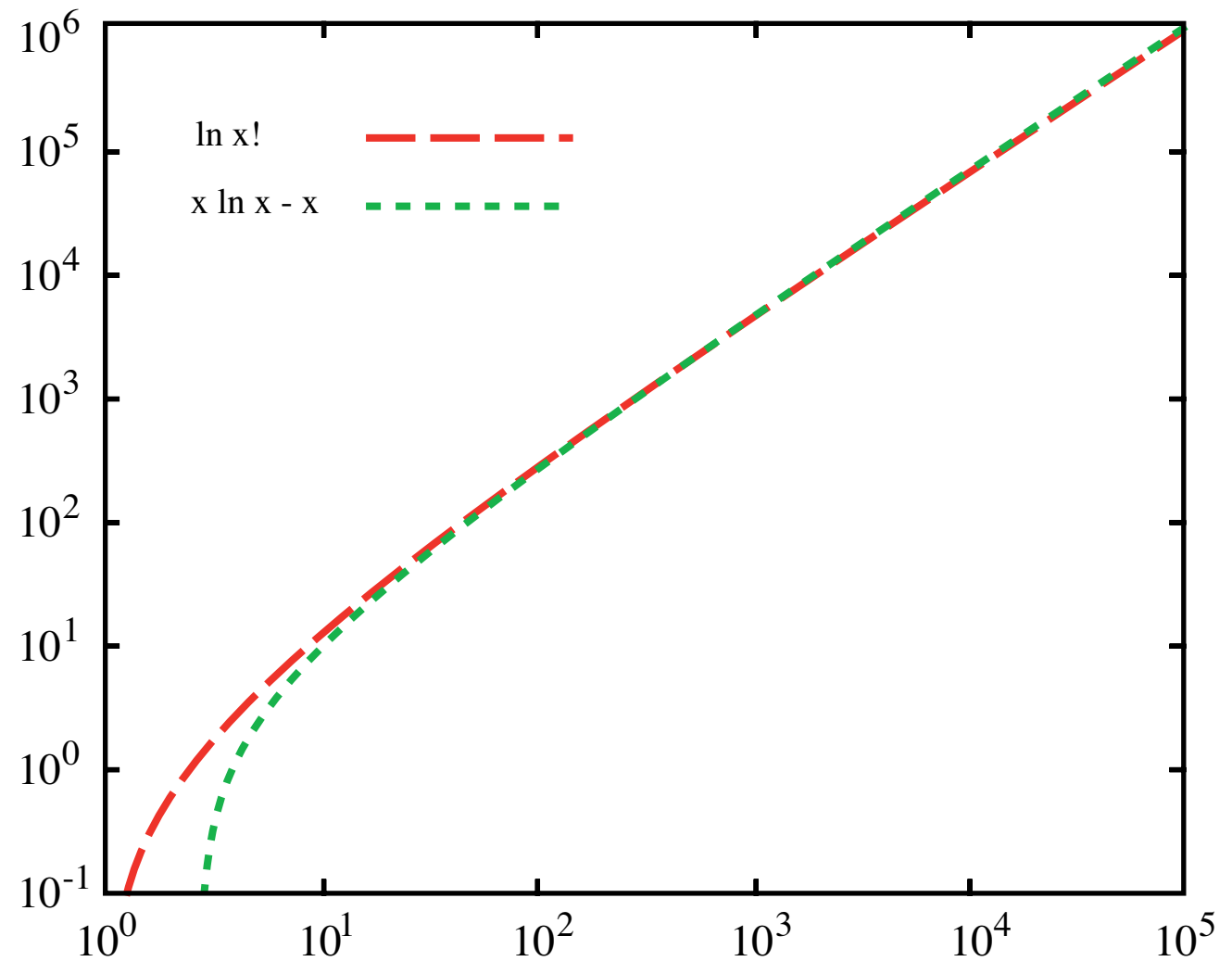

*Fig. 15 - A graphical interpretation of the Stirling approximation*

which, according to our formalism, can be cast in the form of the following (convergent) integral

$$\Delta = \left(\partial_x !\right) \sum_{n=1}^{\infty} (-1)^n e^{n\,x} \,/_{x=0} = \int_0^{\infty} \frac{e^{-t}}{t+1} dt \tag{84b}$$

The research on the above series was inspired to Euler by the solution of the following differential equation

$$\begin{aligned} &x^2 \partial_x y + y = x, \\ &y(0) = 0 \end{aligned} \tag{85}$$

and by the observations on that its series expansion solution

$$y(x) = \sum_{n=1}^{\infty} (-1)^n n!\, x^n \tag{86}$$

does not converge for any, however small, value of x, while its solution in integral form

$$y(x) = x \int_0^{\infty} \frac{e^{-t}}{1 + x\,t} dt \tag{87}$$

is well behaved for any positive *x*.

We will see in the forthcoming section that the previous result can be framed within the more general context of the Borel transform.

## 7. INTEGRAL TRANSFORMS AND EULER-RIEMANN FUNCTIONS

Before concluding our tour we will discuss how the link between Zeta function and particular types of integral transforms (like the Mellin-Laplace Transform) follows from the Euler manipulations.

To this aim we consider the function

$$\zeta(x,p)=\sum_{n=1}^{\infty}\frac{x^n}{n^p} \tag{88}$$

which can also be viewed as of a kind of generalization of the logarithmic function. On account of the identity

$$(x\partial_x)^n x^m = m^n x^m \tag{89}$$

we can write eq. (88) as the following finite expression [14]

$$\zeta(x,p)=(x\partial_x)^{-p}\sum_{n=1}^{\infty}x^n=(x\partial_x)^{-p}\left[\frac{1}{1-x}-1\right] \qquad (90).$$

The function (80) can therefore be written in operational terms as (see eq. (22))

$$\zeta(x,p)=(x\partial_x)^{-p}\frac{x}{1-x} \qquad (91).$$

Furthermore, by exploiting the identities

$$\begin{aligned}
&(x\partial_x)^{-p}=\frac{1}{\Gamma(p)}\int_0^{\infty}e^{-\xi\hat{D}_x}\xi^{p-1}d\xi,\\
&e^{\lambda\hat{D}_x}f(x)=f(e^{\lambda}x),\\
&\hat{D}_x=x\partial_x
\end{aligned} \qquad (92).$$

we end up with the following integral representation

$$\zeta(x,p)=\frac{1}{\Gamma(p)}\int_0^{\infty}\xi^{p-1}\frac{e^{-\xi}}{1-e^{-\xi}x}x\,d\xi \qquad (93).$$

The exponential operator appearing in the first of eqs. (92) is called the dilation operator and its use was originally suggested by Euler himself. It plays a central role in the theory of the E-R function and the associated operator [14]

$$\hat{Z}_p=\frac{1}{\Gamma(p)}\int_0^{\infty}e^{-\xi\hat{D}}\xi^{p-1}d\xi \qquad (94)$$

is the generator of a new transform, in between the Laplace and Mellin transforms [14] and, indeed, its application on a given function yields

$$\hat{Z}_p f(x)=\frac{1}{\Gamma(p)}\int_0^{\infty}f(e^{-\xi}x)\xi^{p-1}d\xi \qquad (95)$$

Furthermore, once acting on the E-R function, provides the following remarkable identity

$$\hat{Z}_p\zeta(x,m)\,|_{x=1}=\zeta(m+p) \qquad (96).$$

Before closing this section, let us discuss the link of the topics we have treated so far with the so called Borel transform.

We consider the series

$$S(x,a_n)=\sum_{n=0}^{\infty}a_n x^n \qquad (97)$$

which can be rearranged as it follows

$$S(x,a_n)=\sum_{n=0}^{\infty}(n!\,a_n)\frac{x^n}{n!}=\int_0^{\infty}e^{-t}\sum_{n=0}^{\infty}a_n\frac{(t\,x)^n}{n!}dt \qquad (98).$$

If

$$\sum_{n=0}^{\infty}\frac{x^n}{n!}a_n=g(x) \qquad (99)$$

we obtain

$$S(x,a_n) = \int_0^\infty e^{-t} g(xt)\,dt \quad (100)$$

thus transforming a (possibly) divergent sum in a (possibly) converging integral. The concepts associated with the Borel transform methods can easily be linked to the discussion we have developed so far. It is easily seen, for example, that

$$S(x,(-1)^n n!) = \int_0^\infty \frac{e^{-t}}{1+xt}\,dt \quad (101)$$

Furthermore the Borel transform of the exponential can be associated with Bessel type function

$$S(x,\frac{(-1)^n}{n!}) = \int_0^\infty e^{-t} C_0(xt)\,dt \quad (102).$$

Where $C_0(x) = \sum_{r=0}^{\infty} \frac{(-x)^r}{(r!)^2}$ is the 0-th order Tricomi function linked to the 0-th order Bessel function by the relation

$$C_0(x) = J_0(2\sqrt{x}) \quad (103).$$

We cannot go deeper into the theory of Borel transform, which would open new speculations on the link between symbolic calculus and Physics.

## 8. CONCLUDING REMARKS

This review paper has dealt with some old research in pure Mathematics, which have provided a decisive impact on the formulation of most recent theories in Physics. The legacy of Euler to modern science has been so wide and profound that we have just grasped on a few outcome of a prodigious mind, which is still a source of inspiration and is influencing the modern thought.

Ramanujan, the famous Indian self thought mathematician, was one of the main interpreter of this heritage and his work on series (divergent or not) opened new scenarios.

Feynman, some time referred as the modern Euler, elaborated new methods to evaluate physical processes, in which the point of view of Euler (and of Ramanujan) becomes crucial.

The renormalization methods are now used in different branch of physics, but some time without the necessary awareness. We consider therefore useful to explain once again that by "renormalizable " theories it is meant theories free from divergences. Within their framework, infinities can be removed, or better they can be controlled and in some sense they can be absorbed into the free parameters of the theory itself. In QED the infinities are absorbed into the quantities reported in Fig. 16.

The meaning of all this is that the sensible quantities are not the bare (without ay interaction) mass, charge… but the "dressed" ones, which contain all the contributions from all the different diagrams.

This fact has further profound consequences, we consider the diagrams reported in Fig. 17, which account for the quantum corrections to the charge (the coupling constant of QED). Denoting by $I(Q)$ the contribution from a single loop, we find [16] (see Fig. 18)

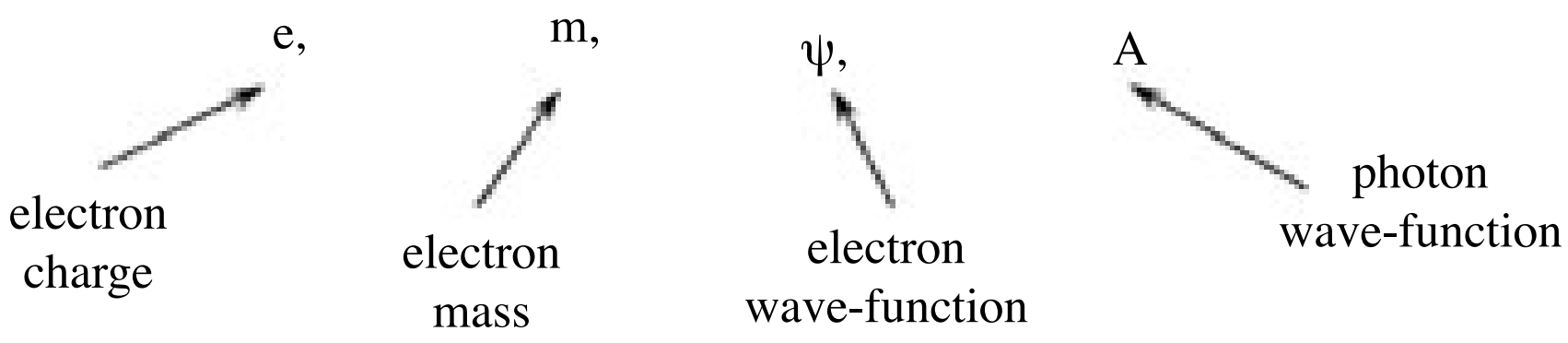

*Fig. 16 - Renormalized quantities of QED*

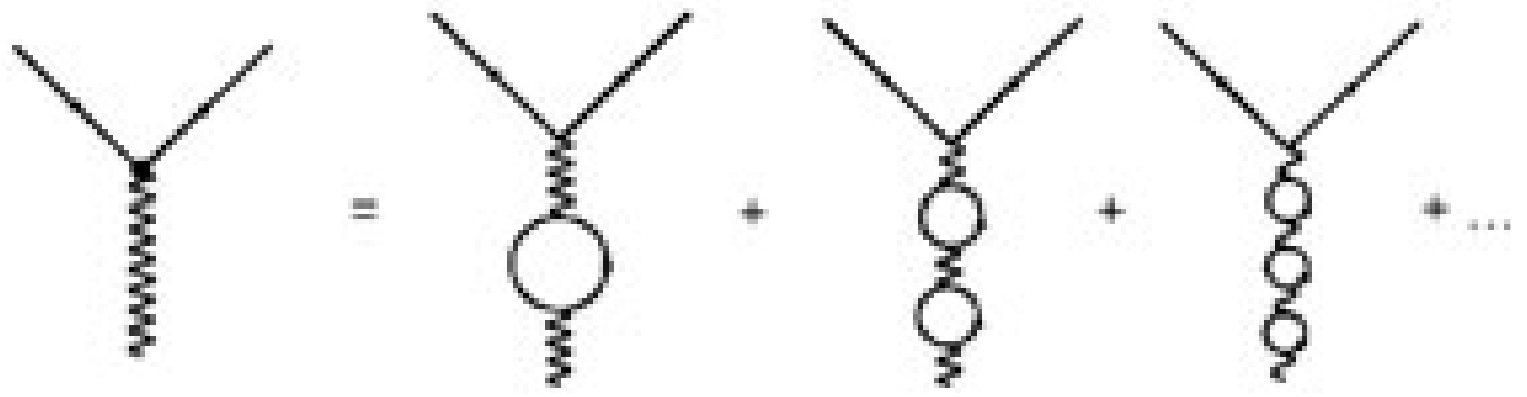

*Fig. 17 - Quantum loop corrections to the electron charge*

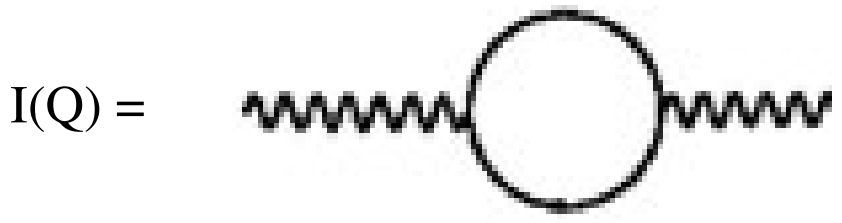

*Fig. 18 - Single loop contribution*

$$e(Q) = e_0 \sum_{n=0}^{\infty} (-1)^n I(Q)^n = \frac{e_0}{1 + I(Q)} \quad (104),$$

It is evident that it can be evaluated at a different energy scale $\mu$ in such a way that

$$e(\mu) = \frac{e(Q)}{1 + I(\mu)} \quad (105)$$

which ensures that charge (or the QED coupling constant) changes with energy, analogous conclusion can be drawn for the QCD. This last observation open consideration associated with the Khallen-Szymanzyk renormalization equation, which is out of the interest of this paper.

We have emphasized the usefulness of the methods associated with ***"the theory of the subtraction of infinities"*** but we have also stressed that, in spite of its success, it does not rely upon solid mathematical grounds. If such a criticism would come from pure mathematicians it could be just considered as the consequence of the perennial dialectic between different ways of conceiving mathematics. If the criticisms come from one of the founders of the theory itself, they should be taken more seriously. We quote therefore Feynman who declared [17].

> **The shell game that we play ... is technically called 'renormalization'. But no matter how clever the word, it is still what I would call a dippy process! Having to resort to such hocus-pocus has prevented us from proving that the theory of quantum electrodynamics is mathematically self-consistent. It's surprising that the theory still hasn't been proved self-consistent one way or the other by now; I suspect that renormalization is not mathematically legitimate.**

We consider appropriate these criticisms and those persistently expressed by Dirac, according to whom [18].

> **Most physicists are very satisfied with the situation. They say: 'Quantum electrodynamics is a good theory and we do not have to worry about it any more.' I must say that I am very dissatisfied with the situation, because this so-called 'good theory' does involve neglecting infinities which appear in its equations, neglecting them in an arbitrary way. This is just not sensible mathematics. Sensible mathematics involves neglecting a quantity when it is small – not neglecting it just because it is infinitely great and you do not want it!**

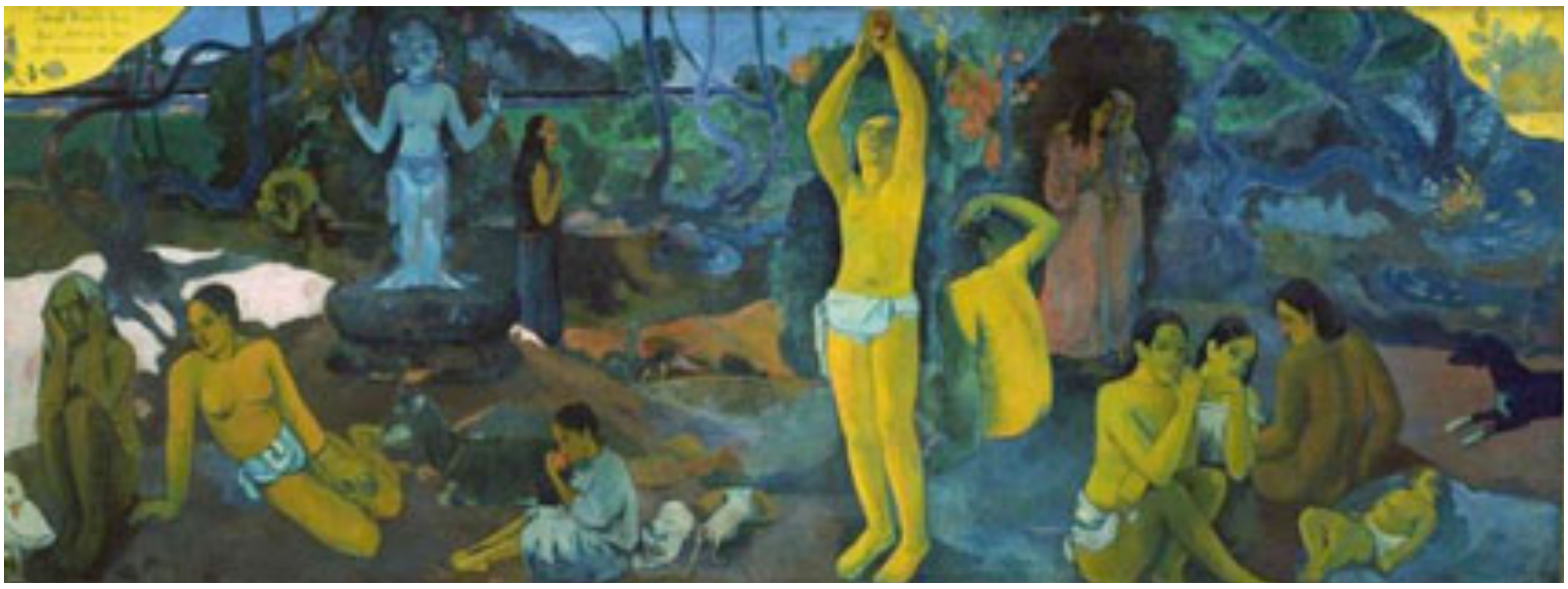

*Fig. 19 - Paul Gauguin : D'où Venons Nous?/ Que Sommes Nous ? Où Allons Nous?*

We are not legitimate to draw any further comment, the best conclusion we may offer to the reader is Fig. 19, reporting the famous Gauguin painting

**Where are we coming from?**
**We are we?**
**Where are we going?**

We would just to add an other question

**Where are we?**

## ACKNOWLEDGMENTS

The Authors recognize that this paper has been stimulated by many pre-coffee morning discussions with A. Doria. The paper is the written version of a seminar, given by one of the Authors (G. D.), at the National Frascati Lab. in 2008, it is therefore a pleasure to thank H. Bilokon for having organized the seminar and have given the chance of discussing these topics with a highly qualified audience. It is finally a great pleasure to thank D. Babusci for a careful revision of the manuscript, for correcting an infinite number of errors and flaws. His criticisms and suggestions have greatly improved the quality of the paper.